
\input phyzzx
\overfullrule=0pt
\date={September 1992}
\Pubnum={PUPT-1341}
\titlepage

\title{ Conformal Turbulence}
\author{A. Polyakov\foot{Supported by NSF Grant PHY90-21984}}
\JHL
\abstract

We describe conformal field theories, correlation functions of which satisfy
equations of the two-dimensional fluid mechanics.  Prediction for the
energy spectrum is given, $E(k) \sim k^{-25/7}$.

\endpage

In this letter I will discuss the turbulent solutions of the equations of
the two-dimensional fluid dynamics.  The term ``turbulent solution" means
here that we will be dealing with the correlation functions of velocities,
vorticities, etc. instead of these quantities themselves.  Therefore, the
implicit assumption in this paper is that in the turbulent flow we have
some well defined stationary probability distribution with respect to which
these correlation functions are defined.

It is well known that, in this
context, the Navier-Stokes equations lead to the chain of relations,
expressing N-point function in terms of N+1 point functions.  The standard
approach to the theory of turbulence is to use some closure hypothesis in
order to get concrete equations for the correlators (see [1] for a review).

Here we take a completely different route.  We will try to satisfy the
above mentioned chain of relations exactly by borrowing sets of
correlation functions, known in the conformal field theory [2].  As a
result we obtain a certain number of exact solutions for the
two-dimensional turbulence.

The basic equations, to be solved, are very
simple.  We have vorticity $\omega$ as our basic fluctuating variable.
Navier-Stokes equations have the form:
$$
\dot\omega + e_{\alpha\beta} \partial_\alpha\psi \partial_\beta
\partial^2\psi = \nu\partial^2\omega \eqno(1)
$$
(Here $\partial_\alpha = \partial /\partial x_\alpha$; $\partial^2 =
{\partial^2\over \partial x^2_\lambda}$.  $\psi$ is a stream function
related to vorticity by $\omega = \partial^2\psi$ and to velocity by
$$
v_\alpha = e_{\alpha\beta} \partial_\beta \psi
$$
$\nu$ is viscosity.)
Now the correlation functions of the type
$$
<\omega (\vec x_1) \cdots \omega (\vec x_N)>
$$
must be time independent, which gives:

$$
<\dot\omega (\vec x_1) \omega(\vec x_2) \cdots> + <\omega(\vec x_1)
\dot\omega (\vec x_2) \cdots > + \cdots \ = \ 0 \eqno(2)
$$

It is understood in (2) that we should express $\dot\omega$ through $\psi$
via eq.(1).  This obvious equation (when being written for the generating
functionals) is called Hopf equation [1].  It expresses the fact that the
probability distribution of our system commutes with the equations of
motion. Since in turbulence we are dealing with the large Reynolds numbers,
the viscous term in (1) can be neglected.  It plays the role of the
ultraviolet cutoff for the theory and will be discussed later.

Let us now present the simplest theory, which satisfies inviscid Hopf
equation (2).  Take the minimal model (2, 5) of conformal field theory (see
[2] for the definitions).  In this case we have two primary fields --- the
unit operator $I$ and the field with the complex dimension (-1/5, -1/5),
which we will identify with $\psi$.  The two point function of $\psi$ in the
conformal theory is:
$$
<\psi(\vec x) \psi(0)> \ =\ - | \vec x |^{4/5} \eqno(3)
$$
We shall postulate that, in order to get physical correlation function from
the conformal one, we have to consider momentum representation of (3):
$$
< \psi (\vec k) \psi (-\vec k) > \ =\ {{\rm const}\over |\vec k|^{2+4/5}}
\eqno(4)
$$
and to transform it back by cutting off the $k$-integral at some infrared
point,
defined by the large scales, $k_{min} \sim 1/r$.  This would give:
$$
<\psi (\vec x) \psi (0)>\ =\ const [R^{4/5} - |\vec x|^{4/5}]. \eqno(5)
$$
We will have in mind similar rule for all cases, namely that the momentum
space correlation functions
$$
<\psi(\vec k_1) \cdots \psi (\vec k_N)>
$$
for $|\vec k_j| >> 1/R$ are R-independent, while in the coordinate space
one should include contributions from the infrared corners.  Thus defined,
physical correlators differ from the conformal ones by the terms, analytic
in $\vec x_i - \vec x_j$.  It should also be understood that conformal
expressions are valid only if
$$
|\vec k| < k_{max} \sim 1/a
$$
The ultraviolet cutoff a determined by the viscosity.  The presence of
the UV cutoff means that we should use careful point-splitting procedure
when defining the products of operators.  In particular the product in the
LHS of (1) should be understood as:
$$
e_{\alpha\beta} \partial_\alpha \psi(\vec x) \partial_\beta \partial^2
\psi(\vec x) =
\overline{\lim_{a\to 0}} e_{\alpha\beta} \partial_\alpha \psi (\vec x +
\vec a /2)
\partial_\beta \partial^2 \psi (\vec x - \vec a /2) \eqno(6)
$$
(where $\overline{lim}$ means directional averaging).
In order to evaluate the RHS of (6) we have to use operator product
expansion, which, for our case, can be written symbolically as
$$
[\psi] \times [\psi] \ =\ [I] + [\psi]
$$
where $[\psi]$ means conformal class of $\psi$, i.e. itself together with
the operators
$$
L_{-n_1} L_{-n_2} \cdots L_{-n_k} \psi ,
$$
$L_{-n}$ being Virasoro generators [2].  More explicitly:
$$\eqalign{
\psi(\vec x +\vec a/2) \psi(\vec x - \vec a/2) &=
|a|^{4/5} \bigl({I + C_1 a^2 L_{-2} I + \cdots}\bigr) \cr
&+
|a|^{2/5} \bigl(
{\psi + C_2 a L_{-1} \psi + a^2 (C_3 L_{-2} + C_4 L_{-1}^2 \psi +
\cdots} \bigr) \cr}\eqno(7)
$$
Performing differentiation of the LHS, we find the leading term:
$$
e_{\alpha\beta} \partial_\alpha \psi \partial_\beta \partial^2 \psi =
{\quad\atop a\to o} const |a|^{2/5} (L_{-2} \bar L^2_{-1} - \bar L_{-2}
L^2_{-1}) \psi \eqno(8)
$$
Antisymmetry of the RHS under complex conjugation is, of course, just the
consequence of the e-tensor at the LHS.  If we rescale
$$
\psi \Rightarrow |a|^{2/5} \psi \eqno(9)
$$
requiring fixed values of velocity at the scale a, the a-dependence of the
RHS disappears.  It is now quite easy to see,  that actually RHS of (8) is
equal to zero.  This happens, because in the (2, 5) model, the field
$\psi$
is degenerate on the second level, satisfying:
$$
(L_{-2}  -  5/2 L^2_{-1}) \psi \ =\ 0 \eqno(10)
$$
Then, because of antisymmetry we get:
$$
\dot\omega \ =\ - e_{\alpha\beta} \partial_\alpha \psi \partial_\beta
\partial^2 \psi \ =\ 0
$$
and the Hopf equation is formally satisfied.  The only thing to check is
that this result is not spoiled by the analytic terms, coming from the
infrared contributions, mentioned above.  Indeed, we dealt with the
strictly conformal theory, disregarding these terms.  In order to check
this let us look at the expression for the operator (8) inserted into some
correlation function in the momentum space.  We have:
$$\eqalign{
<\dot\omega(q) \psi(f_1) \cdots \psi(f_N)> &= \int d^2k [k, q] (k^2 -
(q-k)^2)\cr
&\times <\psi(k) \psi(q-k) \psi(f_1) \cdots \psi(f_N)> .\cr} \eqno(11)
$$
The dangerous region is $k \to 0$.  By using operator products once again
it is easy to see that:
$$
<\psi (k) \psi (q - k) \psi(f_1) \cdots > {\sim \atop {k\to 0}} {1\over
|k|^{2+4/5}} \eqno(12)
$$
and hence there is no infrared divergence in general.

Let us estimate the region in which our solution works.  First of all we
compare kinematical time at the scale $r$:
$$
{1\over t(r)} \sim {v(r)\over r} \sim \psi(r)/r^2  \eqno(13)
$$
with the relaxation time:
$$
{1\over \tau(r)} \sim {\dot\psi (r)\over \psi(r)} \eqno(14)
$$
Using the fact that
$$
\psi (r) \sim \psi (a) ({r\over a})^{2/5}
$$
and eq. (1) we obtain
$$
{1\over \tau (r)} = {\psi (a) a^2\over r^2} \cdot  1/r^2  \eqno(15)
$$
On the other hand, our neglect of the viscosity is justified if
$$
{1\over t(r)} > \nu / r^2 \ ;\ \ \  {\psi (r)\over \nu} > 1 \eqno(16)
$$
And hence
$$
a \sim R \cdot ({\rm Reynolds \ number})^{-5/2} \eqno(17)
$$
It should be realized that the solution we discussed is rather exotic.  It
breaks parity, since the stream functions are pseudoscalars and hence can
exist only in the systems with net rotations or magnetic fields.  Another
unusual feature, which we will discuss now is the fact that as $\nu \to 0$
the fluxes of energy and enstrophy vanish.  This regime requires a subtle
balance between injected rotation, enstrophy and energy.  Still it may be
interesting as a first case in which the Hopf equations are exactly solved
(apart from the trivial Gibbs distributions).

Viscosity and stirring
forces should play the role of the boundary condistions for our solutions.
At present we do not know any rigorous way to implement them, and will have
to make some conjectures, which lead to more generic and physical solutions
than the above.

The standard picture of turbulence involves Kolmogorov's idea of constant
fluxes.  According to Kraichnan [3], in two dimensions the relevant flux
is the flux of enstrophy.  Perhaps the condition of constant flux is
necessary for the matching of the perfect fluid solution with the viscous
region.  Let us try to formulate this condition in our language.  The
enstrophy is a conserving quantity for the perfect fluid given by:
$$
H \ =\ \int \omega^2 (\bar x) d^2 x \eqno(18)
$$

In order to study this quantity in the turbulent regime it is convenient to
introduce the point-split definition (6) (which actually has been exploited
already by Kolmogorov in 1941) and to use the identity
$$
{d\over dt} <\omega (x + r/2)> \omega (x - r/2)> = {\partial \over
\partial r_\alpha} <\omega (x + r/2) \omega (x - r/2)
(v_\alpha (x + r/2)-v_\alpha (x - r/2))> \eqno(19)
$$
The constant $r$-independent enstrophy production will be reached if:
$$
<\omega (x + r/2) \omega (x - r/2) (v_\alpha (x + r/2) -
v_\alpha (x - r /2))> = B r_\alpha \eqno(20)
$$
This is a standard Kolmogorov-like relation.  It has striking analogy with
the axial \ \ (and other) anomalies in quantum field theory.  Indeed, if one
considers massless quantum electrodynamics, and define the axial current as
$$
J_{\mu5} (x) = \lim_{r \to o} < \bar\psi (x + r/2) \gamma_\mu
\gamma_5 \psi (x - r/2)> \eqno(21)
$$
the Dirac equation readily gives:
$$\eqalign{
\partial_\mu J_{\mu 5} (x) &= \lim_{r \to o} (A_\mu (x + r/2) -
A_\mu (x - r/2)) <\bar\psi (x + r/2) \gamma_\mu \gamma_5 \psi
(x - r/2) > \cr
&= {\rm const} \int \epsilon _{\mu\nu\lambda\rho} \partial_\mu A_\nu
\partial_\lambda A_\rho d^4x \not= 0 \cr} \eqno(22)
$$
Physically, chiral charge is produced at the cut-off momenta and
transferred through the momentum space into the physical region.  In
Kolmogorov's case the same happens with enstrophy or with energy.

While eq. (20) is very nice, the next standard step looks quite appalling
to a field theorist.  Namely, one concludes from (20) that  dimensionality
in this formula can be counted as a simple sum of dimensions of $\omega \sim
v/r$ and $v$ thus giving $v \sim r$.
The problem here is that in the field theory dimensions are not additive
and, when we take an operator product, we get at the right-hand side some
defect of dimensions.

We will attempt here to formulate constant flux condition
field-theoretically.  Let us consider some conformal field theory with the
fusion rule:
$$
[\psi] \times [\psi] \ =\ [\phi] + \cdots \eqno(23)
$$
As was argued above we have
$$
\dot\omega \sim (L_{-2} \bar L^2_{-1} - \bar L_{-2} L^2_{-1}) \phi
\eqno(24)
$$
Now, if we rewrite eq. (19) in the form:
$$
<\dot\omega (x + r) \omega (x) > \ =\ r^o
$$
we obtain (recalling that $\omega = \partial^2 \psi$):
$$
(\Delta_\phi + 2) + (\Delta_\psi + 1) = 0 \eqno(25)
$$
$$\Delta_\phi + \Delta_\psi = -3$$
If dimensions had been additive $\Delta_\phi = 2\Delta_\psi$ (which is not
the case of course) we would have obtained Kolmogorov-Kraichnan result
$\Delta_\psi = -1$.
We do not insist now
on the especially slow relaxation, by requiring, as we did in our first
example, that $\phi$ is level two degenerate.  Instead, we need the
condition
$$
\Delta_\phi > 2 \Delta_\psi \eqno(26)
$$
which simply means that RHS of (23) vanishes as $a \to 0$.  In this case we
get inequality
$$
\Delta_\psi < - 1 \eqno(27)
$$
which means that the energy spectrum, given by
$$
E (k) \sim k^{4\Delta_\psi  +1}
$$
is steeper than in Kraichan-Kolmogorov approximation
$$
E (k) \sim k^{-3}.
$$
Of course, it should be remembered when implementing (24) that non-zero
contribution there comes from the regular (infrared) term, while the
conformal term is zero, due to orthogonality between the primary operators.
This is just as well, because otherwise logarithmic terms could be
expected.  Also, the conformal part of the theory is invariant under time
reflection, while the correlator (24) breaks it due to ``infrared leakage"
of time asymmetry.

I do not have complete classification of conformal theories, satisfying
above requirement. There exist an appealing example, however.  Let us
consider minimal models of the type (2, 2 N+1).  From our general
classification [2] it follows that in this case we have a set of N primary
fields,
$$
\psi_1 , \ \psi_2 , \ \cdots \psi_N
$$
with dimensions:
$$
-\Delta_s = {(2N - s) (s - 1)\over 2 (2N + 1)} \eqno(29)
$$
and the algebra:
$$
[\psi_x] \times [\psi_s] = [\psi_{2s-1 } + [\psi_{2s-3}] + \cdots \eqno(30)
$$
for $2s-1 \leq N$, and with replacement $s \Rightarrow 2N + 1 - s$
otherwise.

Constant flux condition (25) takes the form:
$$
\Delta_s + \Delta_{2s-1} = -3 \eqno(31)
$$
Solution to this Diofantian equation is $s=4$ and $N=10$ for which case
$$
\Delta_\psi \ =\ \Delta_4 \ =\ - 8/7 ; \Delta_\phi\ =\ \Delta_7\ =\ -
{13\over 7} \eqno(32)
$$
The energy spectrum for this case is
$$
E(k) \sim 1/k^{3+4/7} \eqno(33)
$$
The experimental value of this exponent seems to lie
between 3 and 4.
Also, this solution preserves parity, because there is no $\psi$-operator
in the product $\psi \times \psi$, and more generally we have symmetry
under parity reflection:
$$
\psi_s \to (-1)^{s+1}  \, \psi_s \eqno(34)
$$

There could be other solutions of these requirements. The major difficulty
of our approach is the somewhat heuristic treatment of the analytic
contributions.  We need to supplement conformal field theory with more
definite rules for their contributions.  Before that, our analyses of the
basic equations is not complete for large enough negative $\Delta$ (our
first example with $\Delta = - 1/5$ seems to be immune to this problem).
Probably, one possible approach is to perturb conformal theory away from
criticality and create a small mass gap which will provide IR-cutoff.
This is left for future work.

The whole picture looks as following.  There are many formal solutions
of the inviscid Hopf equations.  Some of these solutions (for which $\phi$
is level two degenerate) are more stable than the others.  However, only
one or very few solutions satisfy right boundary conditions (in the
momentum space) needed to match them with the viscous region.  All that is
very much like the situation for the laminar flows, where one needs
boundary layer consideration in order to determine the correct inviscid
solution.

Another view on the same problem is the use
of Lagrangian coordinates for the fluid.  In this formulation there is some
similarity to string theory which might be useful.  Namely, it is easy to
see that if we describe the fluid by the fluid coordinate
$$
x_A (\xi_1, \xi_2; t) \ \ A\ =\ 1, 2
$$
$$
x_A (\xi, o) \ =\ \xi_A \eqno(35)
$$
$$
\dot x_A (\xi, t) \ =\ v_A (x (\xi, t), t)
$$
There exist a convenient action for the incompressible fluid
 (in any dimensions):
$$
S = \coeff 12 \int dtd^D\xi  \dot x^2 (\xi, t) + \int p (\xi t)
[{\rm det}
({\partial x_A\over \partial\xi_\alpha}) - 1] d^2\xi dt \eqno(36)
$$
where $p$ is the Lagrange multiplier.

This action has invariance under volume preserving diffeomorphisms, and
should be treated in a stringy way, by introducing ghosts, auxilliary metric
etc.  Perhaps in the steady state (for D=2) we have some effective string
theory describing correlations of $x - s$ or, better to say, of the
``vertex operators":
$$
v_A (x) = \int \dot x_A (\xi) \delta (x- x(\xi)) d^2\xi
$$
Finally, there are many obvious generalizations of the above approach, like
including compressiblity, magnetic fields, passive scalars etc.

I am grateful to E. Siggia, V. Yakhot and A. Zamolodchikov for useful
discussions.  I am also pleased to thank Kurt Gottfried and Andre LeClair
for their kind hospitality at Cornell University, and for providing nice
weather, which helped to complete this work.

\endpage

\leftline{References}
\bigskip
\noindent
[1] Monin, A.S. and Yaglom, A.M., ``Statistical Fluid Mechanics,'' MIT
Press, Cambridge (1975).

\noindent
[2] Belavin, A., Polyakov, A. and Zamolodchicov, A., {\sl Nucl. Phys. \bf
B241}, 33 (1984).

\noindent
[3] Kraichnan, R., {\sl Phys. of Fluids \bf 10}, 1417 (1967).

\end